\documentclass{PoS}
\usepackage{lineno}
\usepackage{listings}
\usepackage{soul}
\title{Toward a Public MAGIC Gamma-Ray Telescope Legacy Data Portal}

\ShortTitle{MAGIC data portal}

\author{
M.~Doro$^a$, 
C.~Nigro$^e$,
E.Prandini$^a$,
A. Tramacere$^b$,
M.~Delfino$^{c,d}$, 
J.~Delgado$^{c,d}$,
E.~do Souto$^c$,
L.~Jouvin$^c$, 
J.~Rico$^c$ for the MAGIC Collaboration\footnote{\texttt{https://magic.mpp.mpg.de/}. For collaboration list see PoS(ICRC2019)1177}
\\
\llap{$^a$} University of Padova and INFN Padova, I-35131 Padova (Italy) \\
\llap{$^b$} Department of Astronomy, University of Geneva, Chemin d'Ecogia 16 - 1290 - Versoix  - Switzerland\\
\llap{$^e$} Deutsches Elektronen-Synchrotron (DESY), D-15738 Zeuthen,Germany\\
\llap{$^c$} Institut de Fisica d'Altes Energies (IFAE), The Barcelona Institute 954
of Science and Technology (BIST), E-08193 Bellaterra (Barcelona), 955 Spain\\
\llap{$^d$} also at Port d'Informaci\'o Cientifica (PIC) E-08193 Bellaterra
(Barcelona) Spain\\
}
\abstract{The MAGIC telescopes are one of the three major IACTs (Imaging Atmospheric Cherenkov Telescopes) for observation of gamma rays in the TeV regime currently operative. MAGIC functions since 2003, and has published data from more than 60 sources, mostly blazars. MAGIC already provides astronomical \texttt{.fits} files with basic final scientific products such as spectral energy distributions, light curves and skymaps from published results. In future, the format of the files can be complemented with further relevant information to the community: a) by including the full multi-wavelength dataset enclosed in a publication, b) providing data in alternative easy-to-use formats such as ASCII or ECSV, which are accessible with other commonly used packages such as \texttt{astropy} or \texttt{gammapy}. Finally, besides high level products, activities have started to provide photon event lists and instrument response functions in a format such that scientists within and outside the community are allowed to perform higher level analysis. A second aim is to provide a full legacy of MAGIC data. This contribution will illustrate the achievements and plans of this activity.}

\FullConference{36th International Cosmic Ray Conference -ICRC2019-\\
		July 24th - August 1st, 2019\\
		Madison, WI, U.S.A.}

\begin{document}

\section{The MAGIC Dataset}
\begin{figure}[h!t]
    \centering
    \includegraphics[width=0.95\linewidth]{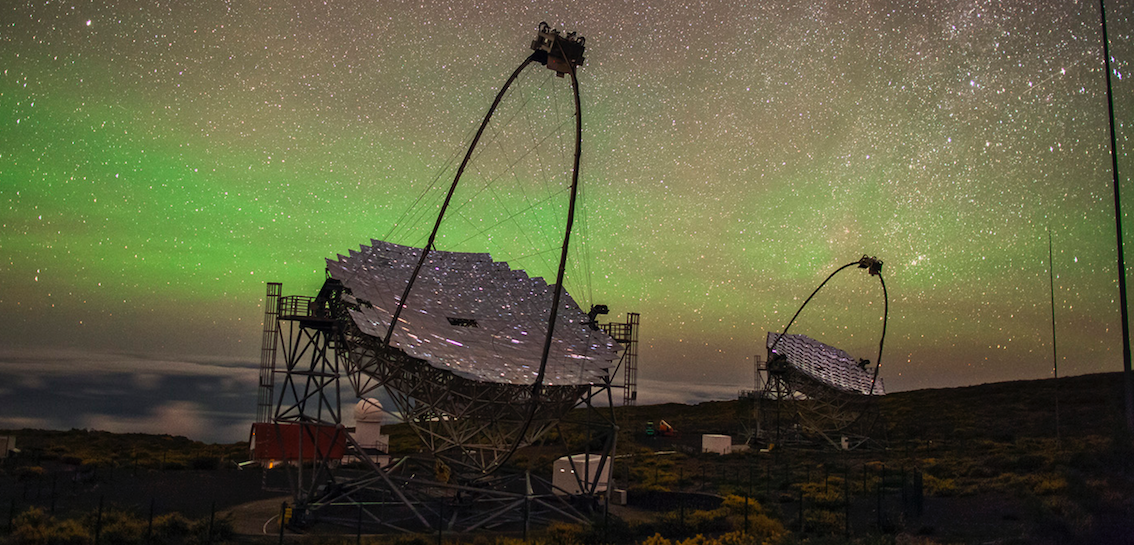}
    \caption{The MAGIC stereo system composed of two 17~m diameter dishes operating simultaneously. Telescopes are located in the Roque de los Muchachos Observatory in the Northern Hemisphere ($28.7^\circ\rm{N}, 17.9^\circ\rm{W}$). Courtesy of D.~Lopes (IAC).}
    \label{fig:magic}
\end{figure}

\begin{figure}[h!t]
    \centering
    \includegraphics[width=0.95\linewidth]{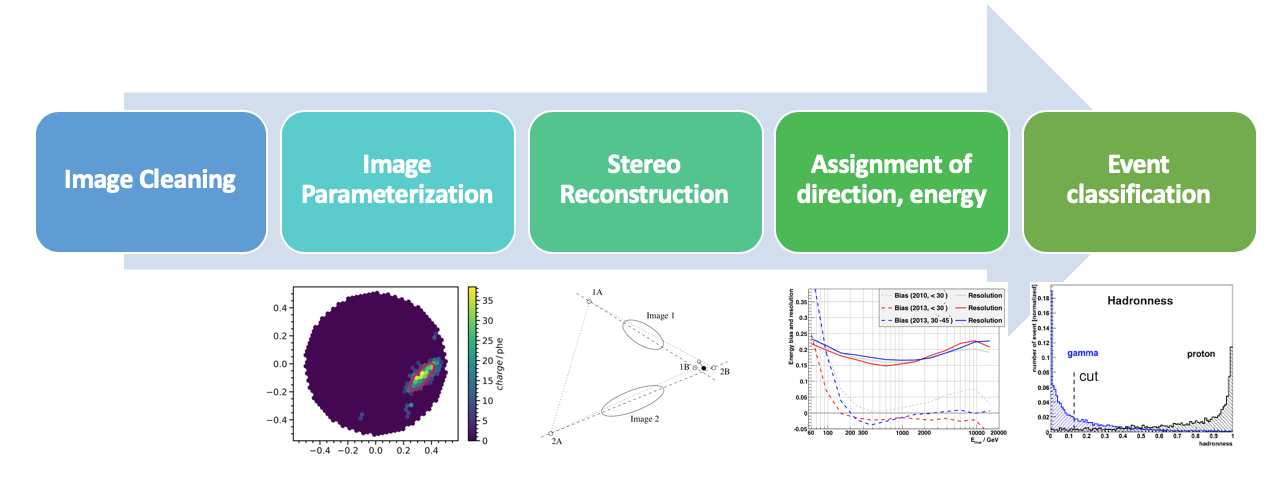}
    \caption{Basic reconstruction steps for raw MAGIC events.}
    \label{fig:dataflow}
\end{figure}

The Major Atmospheric Gamma-ray Imaging Cherenkov (MAGIC) telescopes are a pair of 17~m-diameter telescopes operating in stereo mode and sensitive to cosmic gamma rays with energies between $0.5\div50$~TeV \cite{Aleksic:2014lkm} (see Fig.~\ref{fig:magic}). The first MAGIC telescope operated in standalone mode from 2003 to 2009, when the construction of a second telescope marked the beginning of stereoscopic observations. The imaging technique is based on the detection of Cherenkov light from the charged component of the extended atmospheric showers of particles generated in the high atmosphere ($10-20$~km a.s.l.) by cosmic gamma rays impinging the Earth (as well as any charged cosmic ray). The Cherenkov light propagates as a narrow (few ns wide) front of photons travelling towards the ground that generate an ellipsoidal image on the MAGIC cameras, as shown in Fig.~\ref{fig:dataflow}. 

%\begin{figure}[h!t]
%    \centering
%    \includegraphics[width=0.6\linewidth]{./MAGIC_event_photon.png}
%    \caption{\label{fig:event}A real event from MAGIC reconstructed via %\texttt{ctapipe}~\cite{ctapipe}. The $X,Y$ are camera coordinates and the palette shows the photo-electron intensity. The over-imposed red ellipse represent the parametrized reconstructed event.}
%\end{figure}
%
The Cherenkov light developed by the shower results in a compact cluster of triggered photo-multipliers in the pixelised cameras of both telescopes. If a coincidence of such signals is produced in a few $\sim {\rm ns}$ window the signal of all the pixels of both cameras is read out and stored. Such information constitutes an \textit{event} at the detector level. To extract the physical observables (incoming direction, time, and energy) of the gamma ray producing the event, several algorithms are applied. After subtracting the uniform illumination due to the night sky background (\textit{image cleaning}) the information contained in the aforementioned cluster of triggered pixels is parametrised as an ellipsoidal image \cite{Hillas:1985}, the images of the two cameras are then combined (\textit{stereo reconstruction}) for \textit{the assignment of of direction and energy}, performed with a Random Forest (RF) algorithm and a Monte Carlo look-up table, respectively. The images produced by CR air showers are rejected through a RF classification algorithm \cite{Albert:2007yd} that utilizes Hillas parameters and stereoscopic information returning a single variable ``hadronness'' (close to 0 for gamma-like events). Each step allows for a significant reduction of the data volume. Roughly, of $\sim2\,{\rm TB}$ of raw data per night, $\sim8\,{\rm GB}$ are used for the reduced images (per night, $\sim100\,{\rm MB}$ per observational run), and just $\sim100\,{\rm kB}$ of DL3 data (event lists plus IRFs) per night.

From the above steps one has a list of events characterised by a time of arrival, a direction in the sky and an energy. Several selection cuts are further applied to the data, in energy bins, in order to properly select the signal and background control region.

However, in order to reconstruct a photon flux from these events, one need to know the response of the system: the \emph{Instrument Response Function (IRF)}, that contains the effective area, migration matrices between instrument-derived estimated and Monte Carlo matched coordinates and true energies. IRF components are generated from Monte Carlo simulated events, subject to the same analysis criteria/cuts optimized for the particular observing condition (zenith and azimuth of the observation, the sky quality and brightness, the mirror reflectivity etc...). The event list accompained by IRFs is currently dubbed Data Level 3 (DL3) in the IACT community.
%\cn{analysis cuts optimized to enhance the significance, or somthing like this}
%\cn{diciamo che event lists + IRFs si chiamano DL3 in gergo?}
%
%\begin{figure}[h!t]
%    \centering
%    \includegraphics[width=0.95\linewidth]{./reco.png}
%    \caption{The flux reconstruction flow in MAGIC. See text for details.}
%    \label{fig:reco}
%\end{figure}
%
Ideally, depending the IRF on the observing conditions that continuously change with time, its components should also change for each detected event. As the change is minimal it is usual practice to generate IRFs for the entire observational run. 

It is then clear that besides an event list (time energy direction) ancillary information must be annexed to enable scientists from within and outside the community to reproduce scientific results with MAGIC data. As an intermediate step to releasing the full event list, an activity that is ongoing within MAGIC~\cite{Nigro:2019hqf}, DL3 MAGIC data were produced for the project in Ref.~\cite{Nigro:2019hqf}. MAGIC can share high level products from publications, these comprise e.g. light curves, spectral energy distributions, fits and models, and so on, which can be used for example to build multi-wavelengths data collection. The sharing of these products is extremely simplified with respect to the event list, and it is the main subject of this contribution. However, the possibilities to share the information contained in a publication are also a matter of high debate in our community regarding the technical implementation (format of files, portal server creation, \ldots) and specific information therein. 
While DL3 data have been already produced and shared for projects as the joint-Crab, these proceedings are concerned with sharing the higher level products, also called Data Level 4 (DL4).

Section~\ref{sec:now} reports the current status of data sharing activities;  Section~\ref{sec:future} reports the future plans and Section~\ref{sec:conclusion} closes this report.

\section{Status of data sharing}\label{sec:now} 

The MAGIC Collaboration currently provides a recollection of shared data accessible from the starting page \href{https://magic.mpp.mpg.de/index.php?id=139}{https://magic.mpp.mpg.de/index.php?id=139}. Two products are linked:
\begin{enumerate}
    %\item Virtual observatory: \\ \texttt{magic.mpp.mpg.de/public/public-data/virtual-observatory/}
\item High-level FITS files repository: \href{http://vobs.magic.pic.es/fits/}{http://vobs.magic.pic.es/fits/}
\item Low-level open data repository: \href{http://opendata.magic.pic.es}{http://opendata.magic.pic.es}
\end{enumerate}
%In the Virtual Observatory one can browse MAGIC published observations by inputing sky coordinates and search areas. As an output high-level products such as spectra and lightcurves are provided. 
The High-level FITS repository contains a single \texttt{.fits} file for each publication\footnote{The content of the \texttt{.fits} is specified in a note accessible at the following link: \href{http://vobs.magic.pic.es/fits/mfits/tdas/tdas-fits.pdf}{http://vobs.magic.pic.es/fits/mfits/tdas/tdas-fits.pdf}}, which again contains the high-level products mentioned above as well as skymaps, detection plots and various other pieces of information. The low-level open data contains various lower level events from only a fraction of selected publications.

All of the above represent only an initial step for MAGIC data dissemination, which is going to significantly improve in the near future in several directions:
\begin{description}
\item[Disseminate event list (DL3).] Because of a relative large number of background irreducible events, MAGIC shares candidates event list comprising energy, direction and arrival time. Such list is sometimes dubbed Data Level 3 (DL3) format.
\item[Disseminate High Level products (DL4)] MAGIC produces spectral energy distribution and light curves which are often used in a much larger multi-wavelength framework. Currently \texttt{.fits} files present MAGIC data alone. This data
will be eventually augmented by adding multiwavelength datasets published in MAGIC papers as well as the possibility  of performing queries via API, for example using a \href{}{jupyter notebook}, as in the case of the \href{https://astroquery.readthedocs.io/en/latest/}{astroquery} \cite{astroquery} package.
%Such information is not currently embedded into the \texttt{.fits} files, nor the \texttt{.fits} files are accessible via modern astronomical standard codes such as \texttt{astropy}. 
We aim at improving this situation by providing a novel format of high-level data. This is the subject of the next section.

%\cn{I would rather put a distinction between HIGH-LEVEL PRODUCTS that allow to reproduce the scientific results (DL3) [-> one DL3 file per each observational run] and SCIENTIFIC RESULTS like spectra and lightcurves [-> one result per each publication], those are the concern of this project. I do not think it makes much sense talking about dissemination of single events, unless you are talking about an hypothetic scenario in which the user can manufacture his own event lists having access to all the recorded events and generate IRFs accordingly. What do you do with a single event? :)}

\end{description}

\section{A novel high-level product file}\label{sec:future} 

 One of the main focus of our project is the capability to provide high-level products together with all the metadata regarding the multi-wavelength campaings, that usually are presented in MAGIC papers. 
In general, multi-wavelength data can be retrieved via front-end-based web queries, directly accessing the web service, such as those hosted by the Italian Space Agency (ASI) Science Data Center (SSDC)~\cite{ssdc}; the European Southern Observatory (ESO) Archive Science Portal~\cite{eso}; The Online Data Analysis  \href{https://www.astro.unige.ch/cdci}{ODA}) \cite{ODA} hosted at the University of Geneva; or the Centre de Donnees astronomiques de Strasbourg \cite{cds}. Another approach, more suitable for the automated processing and for processing of large datasets, is the access to the data through  API, as in the case of the ODA-API package \cite{ODA-API} (see for example this \href{https://oda-api.readthedocs.io/en/latest/user_guide/TestAPI.html}{tutorial}), or the 
\href{https://astroquery.readthedocs.io/en/latest/}{astroquery} package.
In this case the query can be performed using few command lines as in the following example

%Spectral points and light curves from astronomical observation are normally found by astronomers via datacenters such as the Italian Space Agency (ASI) Science Data Center (SSDC)~\cite{ssdc}; the European Southern Observatory (ESO) Archive Science Portal~\cite{eso};Aladin?; more?. Another way of accessing general astronomical information is for example through software packages such as \texttt{astropy}~\cite{astropy} that can extract target information via command lines, such as:
%\cn{if you use ''listings'' you have better visualization of code, also syntax highlighting https://en.wikibooks.org/wiki/LaTeX/Source\_Code\_Listings}

\lstset{language=Python} 
%\begin{small}
%\begin{verbatim}
\begin{lstlisting}[frame=single]
from astropy.coordinates import SkyCoord
from astroquery.sdss import SDSS
pos = SkyCoord('0h8m05.63s +14d50m23.3s', frame='icrs')
xid = SDSS.query_region(pos, spectro=True)
print(xid)
\end{lstlisting}
%\end{verbatim}
%\end{small}
%\lstset{language=bash}

There are also some TeV-specific web portal for catalogs of sources such as the TeVCAT~\cite{tevcat}, the TeGeV catalogues~\cite{tegev}. SSDC data for example are exported in ASCII format in the following format:

\begin{small}
%\begin{verbatim}
\begin{lstlisting}[frame=single,language=sh]
# RA = 03 19 48.10(49.950417)
# Dec = +41 30 43.00(41.511944)
# Redshift = 0.0
# NH = 1.36E21 
# BAT60AGN (15 - 55 keV) (id = 54)
# LogFreq (Hz)Bin Nufnu (erg cm^-2 s^-1) Nufnu_error  TStart TStop
18.841672     0.000000    -10.337110   0.011196    53430.0    55256.0
# BAT54MCAT (15-50keV) (id = 53)
# LogFreq (Hz)Bin Nufnu (erg cm^-2 s^-1) Nufnu_error  TStart TStop
18.820989     0.000000    -10.437238   0.022993    53340.5    54525.5
# 2MASS (id = 7)
# LogFreq (Hz)Bin Nufnu (erg cm^-2 s^-1) Nufnu_error  TStart TStop
14.385428     0.000000    -10.443387   0.030683 0.0   0.0
14.256477     0.000000    -10.477122   0.032840 0.0   0.0
14.143015     0.000000    -10.521102   0.025126 0.0   0.0
\end{lstlisting}
%\end{verbatim}
\end{small}

Since  MAGIC  data need some assumptions and ancillary information, the final high-level products cannot be distributed as a single object, as in the case of the  SDSS example above, where the query is returning a single \texttt{astropy} Table. In order to explain the details, we take as an example one specific MAGIC publication AA617(2018)A91~\cite{Ansoldi:2018sqg} , related to the study of the flaring activity of the active galaxy NGC~1275 in 2016-2017.
\begin{figure}[h!t]
    \centering
    \includegraphics[height=5cm]{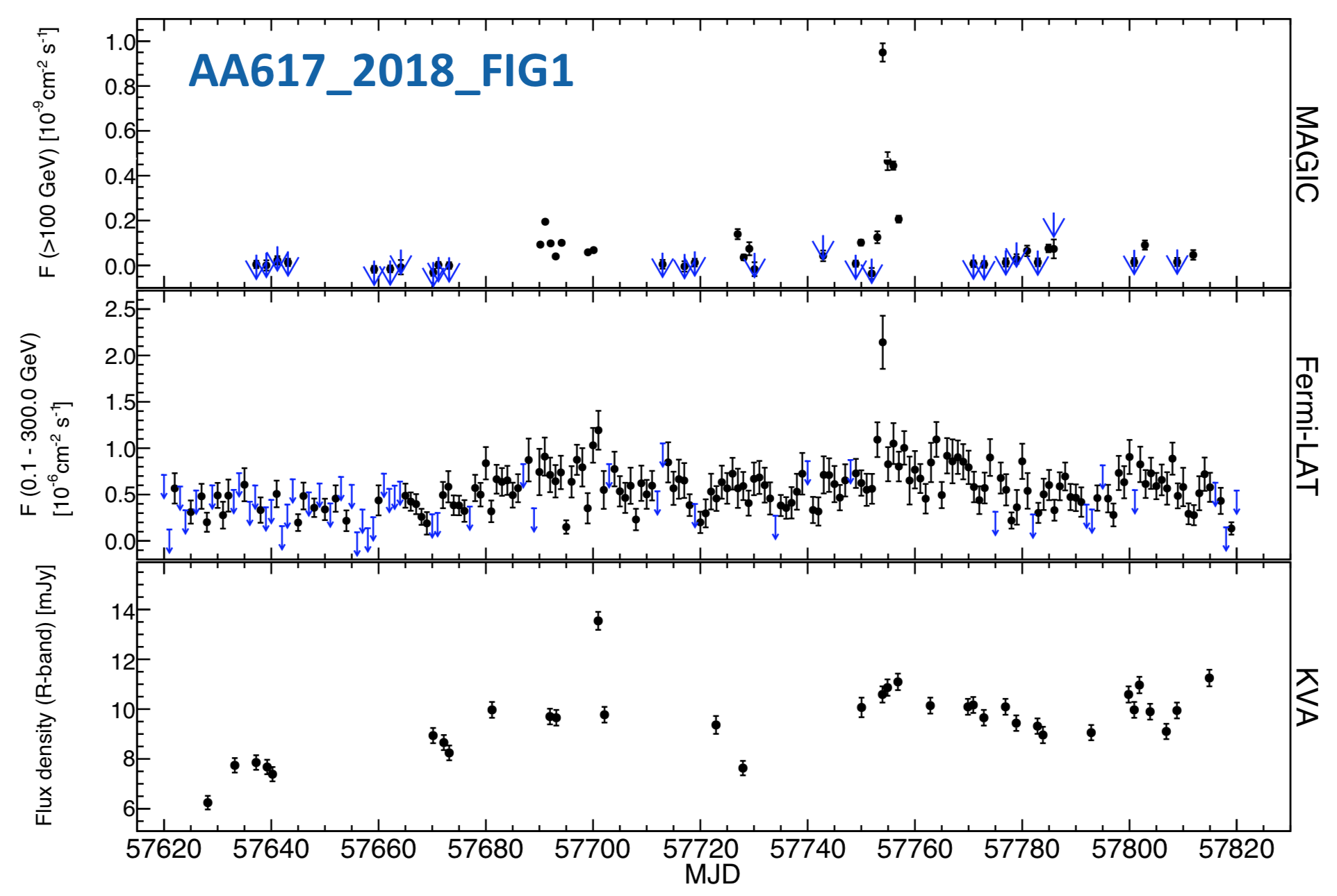}
    \includegraphics[height=5.2cm]{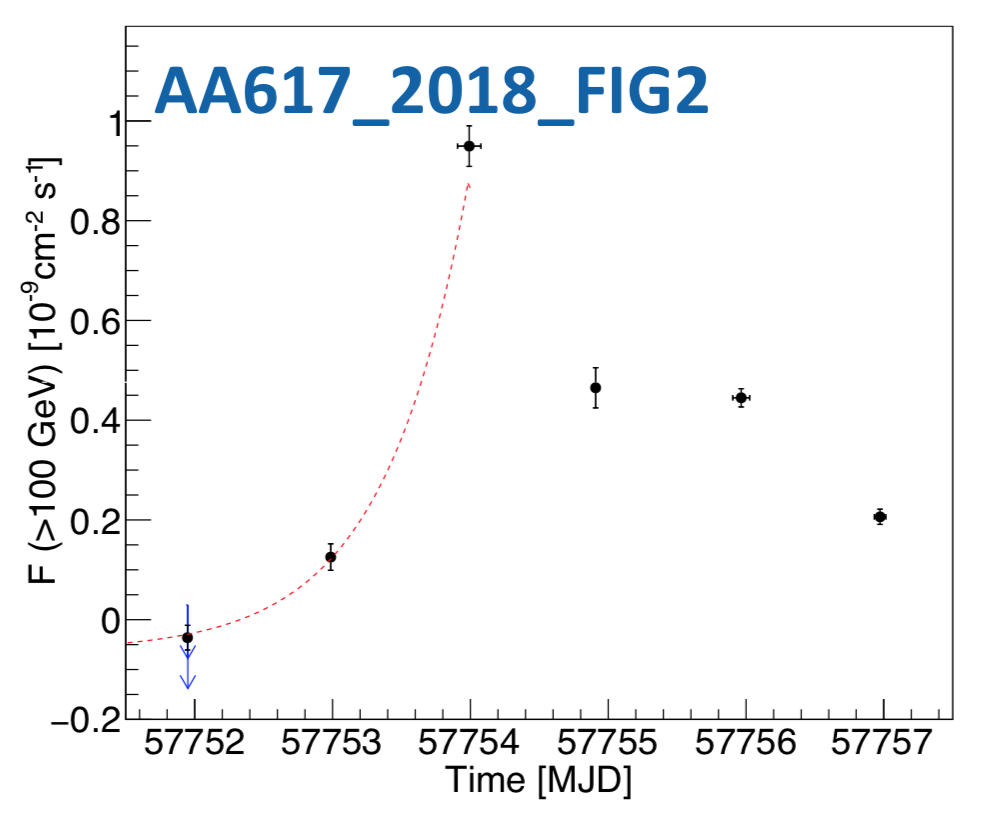}
    \includegraphics[width=0.45\linewidth]{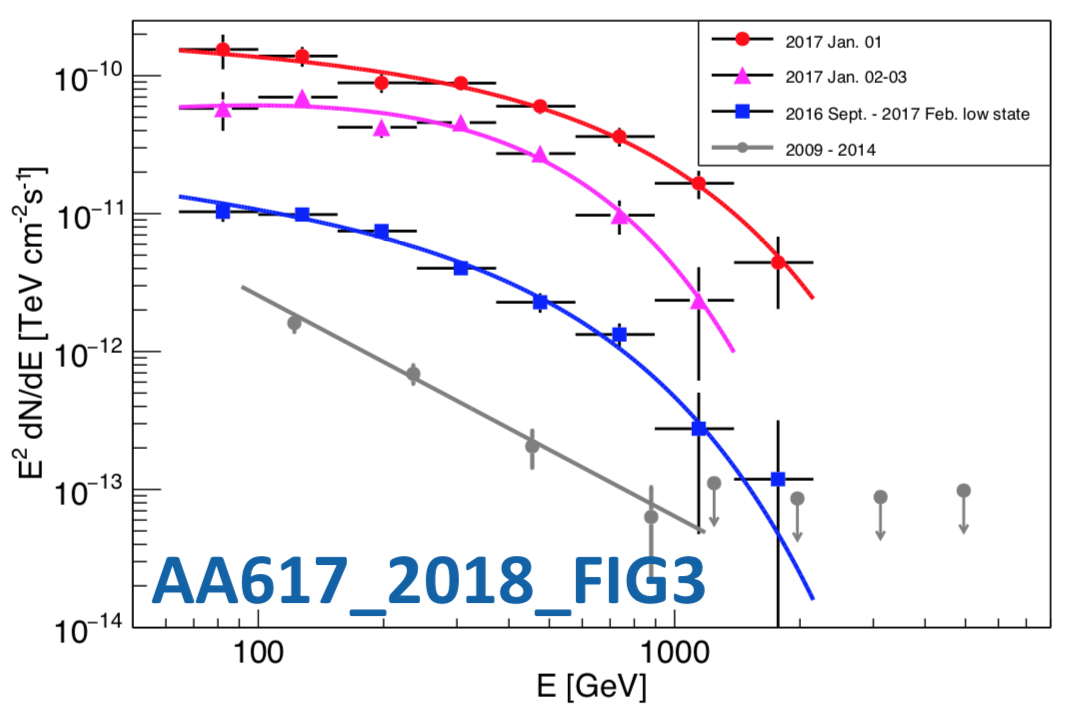}
    \includegraphics[width=0.45\linewidth]{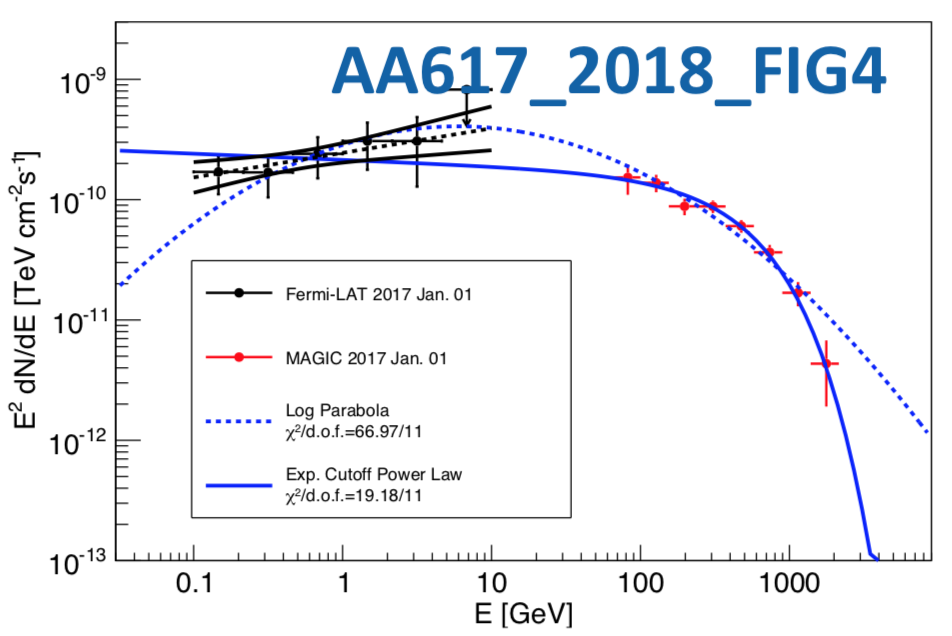}
\caption{\label{fig:paper}Figures of Ref.\cite{Ansoldi:2018sqg}.}
\end{figure}

In Fig.~\ref{fig:paper} we report four of five figures of  AA617(2018)A91: Fig.~\ref{fig:paper}A a MWL LC, Fig.~\ref{fig:paper}B a zoom on this LC, Fig.~\ref{fig:paper}C several MAGIC SEDs for different states, Fig.~\ref{fig:paper}D a MWL SED. The fifth figure in the paper investigates the correlation between wavelengths and it is not of interest here. Of the above four, cleary the second is a zoom of the first and not of interest. But the remaining three are of interest. Fig.~\ref{fig:paper}A reports data of MAGIC, {\it Fermi}/LAT and KVA, Fig.~\ref{fig:paper}D of MAGIC and {\it Fermi}/LAT. In the following paragraphs, we try to investigate the formats of the above mentioned information that the MAGIC Collaboration may distribute to the community.

The format in which high-level data  should be distributed in still a matter of debate within the TeV community. There is a forum for discussion in \texttt{github}~\cite{format}, in which specific tag and fields are discussed, however, a consensus is not yet obtained. Our proposal is to store the metadata regarding each paper into a \texttt{.yaml} file, and the spectral/temporal information into \texttt{astropy} tables, using the extended \texttt{ecsv} format. Both these formats, are largely used in the astronomical community, and easily convertible (eg. \texttt{.yaml} to \texttt{.json}, and \texttt{.ecsv} to \texttt{.fits}) using well established libraries. 

\newpage
As an example, for AA617(2018)A91 we propose to have a \texttt{.yaml} file in the following format:
%
%\newpage
\begin{small}
%\begin{verbatim}
\begin{lstlisting}[frame=single,language=sh]
Filename:: magic_18f.yaml
File_info:
  Fdate = 20190315                             # File creation date
  Fvers = 1                                    # File version 
  Fgen = Michele Doro, michele.doro@unipd.it   # Owner
  Fmail = magic_sapo@mpp.mpg.de                # Contact
  Flink = XXXX                                 # Link
Paper_info:
  Pref: Astron.Astrophys. 617 (2018) A91 
  Pdoi: https://doi.org/10.1051/0004-6361/201832895
  Parxiv: http://arxiv.org/abs/arXiv:1806.01559
  Pcoll: magic                                 # Main collaboration
  Pcauthor: XX, YY                            # Main authors of papers
  Pads:  2018A&A...617A..91M                   # ADS paper tag
  Pinspire: Ansoldi:2018sqg                    # inspire paper tag
Targets in file 
  Tpname: NGC1275              # Main target name
  Taname: 3C84 BZUJ0319+4130 1H0316+413 4C+41.07 ... # Other target names
File list MAGIC:                               # List of MAGIC products
  magic_18f_lc1_fig1.ecsv
  magic_18f_lc2_fig1.ecsv 
  magic_18f_sed1_fig3.ecsv 
  magic_18f_sed2_fig3.ecsv 
  magic_18f_sed3_fig3.ecsv 
  magic_18f_sed1_fig4.ecsv 
File list MWL:                                  # List of MWL products
  magic_18f_lc1_fig1_lat.ecsv
  magic_18f_lc2_fig1_lat.ecsv 
  magic_18f_sed1_fig4_lat.ecsv 
File on demands (available on request to Fmail) # Further products
  magic_18f_sed1_fig3_fit.ecsv 
  magic_18f_sed2_fig3_fit.ecsv 
  magic_18f_sed3_fig3_fit.ecsv 
  magic_18f_sed1_fig4_fit.ecsv 
  magic_18f_sed1_fig4_model.ecsv 
  magic_18f_sed2_fig4_model.ecsv 
Comments: None
\end{lstlisting}
%\end{verbatim}
\end{small}

\noindent
For example \texttt{magic\_18f\_sed1\_fig3.ecsv} would look like:

\begin{small}
%\begin{verbatim}
\begin{lstlisting}[frame=single,language=sh]
# %ECSV 0.9
# ---
# datatype:
# - {name:en, unit:GeV,  Energy}
# - {name:en_wlo, unit:GeV,  Energy bin width low}
# - {name:en_wup, unit:GeV,  Energy bin width up}
# - {name:nufnu, unit:TeV cm-2 s-1 , ,  Dif. ph. flux at en}
# - {name:nufnu_elo , unit: TeV cm-2 s-1,  Low Stat err. on nufnu}
# - {name:nufnu_eup , unit: TeV cm-2 s-1,  Up Stat err. on nufnu}
# - {name:tstart, unit: mjd,  MJD start}
# - {name:tstop, unit: mjd,  MJD stop}
# - {name:texpo, unit: h,  Observation time}
# - {name:comments, unit: latex,  Comments}
# meta: !!omap
# - {Filename: magic_19e_sed_fig1_target01.ecsv}
# - {Source:   TXS0210515}
# - {Comments: }
# schema : astropy

en  en_wlo en_wup nufnu nufnu_elo nufnu_eup tstart tstop texpo comments
0.1567 0.0308 0.3561 1.529e-10 1.429e-10 1.729e-10 57637.1 57811.9 63
0.2484 0.0489 0.5646 4.719e-11 4.619e-11 4.819e-11 57637.1 57811.9 63
0.3937 0.0775 0.8949 1.459e-11 1.359e-11 1.659e-11 57637.1 57811.9 63
0.6239 0.1226 1.4182 4.609e-12 4.509e-12 4.809e-12 57637.1 57811.9 63
0.9888 0.1944 2.2477 1.139e-12 1.039e-12 1.239e-12 57637.1 57811.9 63
1.5670 0.3081 3.5623 3.839e-13 3.739e-13 4.139e-13 57637.1 57811.9 63
\end{lstlisting}
%\end{verbatim}
\end{small}

The above example file is significantly based on the format defined for the \texttt{gamma-cat} project~\cite{gammacat}.  This project aims at collecting all publications from current IACTs, and at being considered a reference for future projects. Our proposal provides a wider dataset that can be straightforwardly exported into the \texttt{gamma-cat} standards. 

There are several possibilities to store such files. Considering we would like to provide an API, our server could be based in one of the MAGIC Collaboration institutes, and mirrored elsewhere. The space requirement would be minor. The most critical part of the project is that of iterating all published papers and transfer all the high-level products into the proposed format. However, for novel published papers this would not constitute a problem.

\section{Discussion and conclusions}\label{sec:conclusion} 
We believe that providing high-level product data from MAGIC publications may be of interest to a wider astronomical community. We want to facilitate the access to these data by building an easily accessible dataset for all MAGIC published paper as well as papers to come. This paper catalogue will be released in 2019 and will be filled constantly.

Besides this effort, the MAGIC Collaboration is also working on the dissemination of the data at DL3-level, with high level data products relative to individual observations. A working group is active in MAGIC and results are expected soon. 

\paragraph{Acknowledgement}
This proposal comes from discussion with several people, specially those attending the 
ASTERICS-OBELICS PyGamma19 meeting in Heidelberg \\ \href{https://indico.cern.ch/event/783425/overview}{https://indico.cern.ch/event/783425/overview} in particular C.~Boisson, C.~Deil, G.~Maier and R.~Zanin. This project has received funding from the European Union's Horizon2020 research and innovation programme under the Marie Sklodowska-Curie grant agreement no 664931. The MAGIC Collaboration acknowledges support of institutes as in \\
\href{https://magic.mpp.mpg.de/acknowledgments\_ICRC2019/}{https://magic.mpp.mpg.de/acknowledgments\_ICRC2019/}.

\end{document}